\documentclass{article}

\usepackage{arxiv}
\usepackage[T1]{fontenc}    
\usepackage{hyperref}       
\usepackage{url}            
\usepackage{booktabs}       
\usepackage{amsfonts}       
\usepackage{amsmath}
\usepackage{nicefrac}       
\usepackage{microtype}      
\usepackage{graphicx}
\usepackage[numbers]{natbib}
\usepackage{doi}

\usepackage{float}
\usepackage{outlines}
\usepackage{subfigure}
\usepackage{multirow, booktabs}

\usepackage{xcolor}

\newcommand{\fontDiscrete}{\mathcal}
\newcommand{\dummySymb}{v}

\newcommand{\paramSymb}{\mu}
\newcommand{\solSymb}{x}
\newcommand{\sol}{\boldsymbol \solSymb}

\newcommand{\param}{\boldsymbol \paramSymb}

\newcommand{\solArg}[1]{\sol_{#1}}

\newcommand{\RR}[1]{\ensuremath{\mathbb{R}^{ #1 }}}

\newcommand{\nbig}{N}
\newcommand{\spaceSymb}{s}
\newcommand{\nspacedof}{\nbig_\spaceSymb}

\newcommand{\nsmall}{n}
\newcommand{\nparam}{\nsmall_\mu}
\newcommand{\spatialSubspace}{\fontDiscrete S}
\newcommand{\defeq}{:=}
\newcommand{\scaledDecoder}{\boldsymbol{g}}
\newcommand{\scaledEncoder}{\boldsymbol{h}}
\newcommand{\dummy}{\boldsymbol {\dummySymb}}
\newcommand{\reddummy}{\hat{\dummy}}
\newcommand{\nbasisspace}{{\nsmall_\spaceSymb}}
\newcommand{\solapprox}{\tilde\sol}
\newcommand{\redsolapprox}{\hat\sol}
\newcommand{\redsolapproxArg}[1]{\redsolapprox_{#1}}
\newcommand{\basismatspaceSymb}{\Phi}
\newcommand{\basismatspace}{\boldsymbol{\basismatspaceSymb}}

\newcommand{\snapshotSymb}{X}
\newcommand{\snapshots}{\boldsymbol \snapshotSymb}
\newcommand{\bmat}[1]{\begin{bmatrix}#1\end{bmatrix}}
\newcommand{\timeSymb}{t}
\newcommand{\ntimedof}{{\nbig_\timeSymb}}

\newcommand{\ndof}{\nspacedof}

\newcommand{\argmin}[1]{\underset{#1}{\text{argmin}}}
\newcommand{\samplematSymb}{Z}
\newcommand{\samplematNT}{\boldsymbol \samplematSymb}
\newcommand{\samplemat}{\samplematNT^T}

\newcommand{\resSymb}{r}
\newcommand{\nbasisres}{\nsmall_\resSymb}

\title{Real-Time Reconstruction of Ground Motion During Small Magnitude Earthquakes: A Pilot Study}

\author{Youngkyu Kim\thanks{Corresponding authors} \\
	Intelligence and Interaction Research Center\\
	Korea Institute of Science and Technology\\
	Seongbuk-gu, Seoul 02792, Republic of Korea \\
	\texttt{youngkyu.kim@kist.re.kr} \\
	\And
	Qingkai Kong \\
	Atmospheric, Earth, \& Energy Division \\
	Lawrence Livermore National Laboratory \\
	Livermore, CA 94550, USA\\
	\texttt{kong11@llnl.gov} \\
	\AND
	Youngsoo Choi \\
	Center for Applied Scientific Computing \\
        Lawrence Livermore National Laboratory \\
	Livermore, CA 94550, USA\\
	\texttt{choi15@llnl.gov} \\
	\AND
	Arben Pitarka \\
        Atmospheric, Earth, \& Energy Division \\
	Lawrence Livermore National Laboratory\\
	Livermore, CA 94550, USA\\
	\texttt{pitarka1@llnl.gov} \\
        \And
	Byounghyun Yoo\footnotemark[1] \\
	Intelligence and Interaction Research Center  \\
        Korea Institute of Science and Technology \\
        AI-Robotics, KIST School\\
        Korea National University of Science and Technology\\
	Seongbuk-gu, Seoul 02792, Republic of Korea\\
	\texttt{yoo@byoo.org} \\
}
\date{}

\renewcommand{\shorttitle}{\textit{arXiv} Template}

\begin{document}
\maketitle

\begin{abstract}
This study presents a pilot investigation into a novel method for reconstructing real-time ground motion during small magnitude earthquakes (M < 4.5), removing the need for computationally expensive source characterization and simulation processes to assess ground shaking. Small magnitude earthquakes, which occur frequently and can be modeled as point sources, provide ideal conditions for evaluating real-time reconstruction methods. 
Utilizing sparse observation data, the method applies the Gappy Auto-Encoder (Gappy AE) algorithm for efficient field data reconstruction. This is the first study to apply the Gappy AE algorithm to earthquake ground motion reconstruction. 
Numerical experiments conducted with SW4 simulations demonstrate the method's accuracy and speed across varying seismic scenarios. The reconstruction performance is further validated using real seismic data from the Berkeley area in California, USA, demonstrating the potential for practical application of real-time earthquake data reconstruction using Gappy AE. As a pilot investigation, it lays the groundwork for future applications to larger and more complex seismic events.
\end{abstract}


\section{Introduction}
Earthquake-imposed risks require rapid and accurate assessment of ground motion to enable effective disaster management. Traditional physics-based methods for ground motion simulation, such as SW4 \cite{sw4-v3.0, Sjogreen2012sw4} and SPECFEM \cite{Dimitri2000, Komatitsch1999}, provide accurate ground motion predictions. However, these methods require significant computational resources and time due to source characterization, which focuses on reconstructing seismic activity from fault parameters, thereby restricting their real-time applicability \cite{McCallen2020, McCallen2021,Graves2014,Maufroy2015,10.1002/eqe.3377,10.1785/0220180261,Kuang2021}. Recently, there are also machine learning-based approaches for ground motion simulations as surrogate models for fast seismic simulations \cite{10.1146/annurev-earth-071822-100323, 10.1785/0220180259, 10.5194/se-11-1527-2020, 10.2172/2001189, 10.1109/tgrs.2023.3264210, 10.1093/gji/ggae342}, but most of the studies are not designed in real-time. Meanwhile, Early Earthquake Warning Systems (EEWS) are essential for providing rapid alerts by predicting seismic intensity based on the initial onset of the P-waves \cite{10.1002/eqe.4029, 10.1080/13632469.2020.1826371, 10.1785/0220200006, 10459332, 10.1016/j.earscirev.2020.103184, 10.1146/annurev-earth-053018-060457}. Despite their importance, EEWS are limited in their ability to deliver detailed spatial and temporal ground motion histories. These systems lack the capability to provide location-specific ground motion predictions over time, emphasizing the need for a fast and spatially detailed reconstruction methodology. This presents a significant challenge in earthquake hazards management and mitigation. Thus there is a need to develop accurate and detailed ground motion predictions in real-time to enable more effective response during seismic events.

This pilot study focuses on small magnitude earthquakes (M < 4.5) modeled as point sources to explore the feasibility of real-time ground motion reconstruction. Small magnitude earthquakes occur frequently and typically exhibit simpler rupture characteristics, allowing them to be approximated as point sources. These features make them particularly well-suited for testing and validating a novel reconstruction technique. The objectives are (1) to propose a real-time seismic ground motion reconstruction method based on sparse observation data, and (2) to demonstrate its feasibility using the Gappy AE algorithm, bypassing the need for high computational cost direct simulations. If earthquake ground motions in surrounding areas can be reconstructed in real-time using sparse measurement from existing seismic stations, it can serve as an effective tool for compensating for the non-uniform and incomplete configuration of the seismic network. Additionally, by recording real-time ground motion, this approach can be quickly utilized after an earthquake to assess affected areas or evaluate at-risk buildings.

Existing studies related to the reconstruction of earthquake data can be summarized as follows. Seismic data acquisition often faces challenges due to economic, environmental, and technical constraints, resulting in irregular and incomplete data collection \cite{Zhang2023, Canning1996}. This compromises the accuracy of ground shaking estimation, subsurface structure interpretation and negatively impacts seismic hazards and geological structure analysis. To address these issues, various data reconstruction and interpolation methods have been developed. Traditional approaches include wave equation-based reconstruction, predictive filtering, and sparse transform-based methods \cite{Kuijpers2021, Wang2019_1}. Each of these methods has varying applicability depending on the characteristics of the data. Recently, novel approaches such as the Low-Dimensional Manifold Model (LDMM) have also been proposed \cite{Lan2022}. Deep learning-based methods have gained significant attention, with various architectures like U-Net, Convolutional Neural Networks (CNNs), and Generative Adversarial Networks (GANs) being applied \cite{wang2019deep,chai2020deep,wang2020seismic,Goyes2024}. These methods can learn complex patterns and enable high-quality data reconstruction, but they often require large training datasets and may have limitations in generalization capabilities. Recent efforts focus on improving deep learning model performance through techniques such as self-supervised learning and transfer learning \cite{Zhu2023,Zhu2024}. Additionally, hybrid methods combining compressed sensing and deep learning, as well as approaches using diffusion models, have been proposed \cite{Wu2023,deng2024}. The low-rank structure of Hankel matrices formed from seismic data has been exploited for effective denoising and interpolation \cite{Wang2019_1, Chen2016,Wang10103709}. This property has been leveraged in both traditional and machine learning-based approaches.

Unlike the aforementioned studies, we adopt the Gappy AE algorithm \cite{KIM2024116978}. This method utilizes an autoencoder to learn the nonlinear mapping between seismic data and the latent space. By solving an error minimization problem between the measured values and the predictions of the decoder, it enables real-time computation of coordinates in the latent space to reconstruct surrounding data. The key contributions of this pilot study are as follows: (1) the first application of the Gappy AE algorithm to seismic data, (2) real-time data reconstruction using less than 2\% of sparse measurement data, (3) a comparative analysis of reconstruction performance based on different sampling algorithms, and (4) validation of reconstruction performance using real small magnitude earthquake data from the Berkeley area in California, USA.

This paper is organized as follows. First, we introduce the Gappy AE algorithm and evaluate its performance under different sampling techniques using simulation data. Then, we assess the performance of the Gappy AE algorithm on real field data and provide an in-depth discussion of the analysis. Finally, we conclude this study.

\section{Gappy AE}\label{sec:Gappy AE}
When dealing with gappy data, a commonly used traditional method for data reconstruction is the Gappy Proper Orthogonal Decomposition (Gappy POD) approach. This method utilizes Proper Orthogonal Decomposition (POD) to estimate the complete dataset \cite{everson1995karhunen, bui2004aerodynamic, willcox2006unsteady}. However, Gappy POD operates within a linear subspace, limiting its effectiveness in restoring solutions with large Kolmogorov N-width characteristics \cite{greif2019decay, kim2022fast, KIM2024116978}. Seismic data, which exhibit large Kolmogorov N-width, are particularly ill-suited for this approach. In contrast, the Gappy Auto-Encoder (Gappy AE) method, which employs a nonlinear manifold using an Auto-Encoder (AE)—an unsupervised learning model in artificial intelligence—demonstrates superior data representation capabilities, particularly for data with high advection characteristics and seismic waves. The Auto-Encoder's ability to learn complex nonlinear relationships results in lower projection errors and improved reconstruction quality. This method allows for reconstruction without being constrained by the properties of the data space \cite{KIM2024116978}. Thus, this study employs the Gappy AE method for seismic data reconstruction, following the explanation provided in \cite{KIM2024116978} to ensure the self-containment of this paper.

Gappy AE is designed to reconstruct missing values in sparse data by leveraging the architecture of an Auto-Encoder. Gappy AE utilizes nonlinear manifold learning, making it particularly effective for handling data with complex characteristics. The method comprises an encoder and a decoder \cite{KIM2024116978}. The encoder compresses high-dimensional data into a compact latent space representation, capturing essential features while reducing redundancy. The decoder then reconstructs the data from this latent space back to their original form. The nonlinear nature of the manifold enables the Auto-Encoder to model intricate data distributions and relationships beyond the capabilities of linear methods.

The reconstruction process begins by estimating initial latent space coordinates by inputting an initial value into the encoder. This step is followed by solving an optimization problem that minimizes the error between the known sparse measurements (seismic sensor measurements) and the reconstructed values produced by the decoder. The optimization is solved iteratively using the Gauss-Newton method, ensuring convergence to an accurate reconstruction. The computed latent space coordinates are then utilized to generate the final reconstructed data.

One key feature of Gappy AE is its use of a specialized sparse connection structure within the neural network. These sparse connections enhance computational efficiency and improve the network's generalization ability when dealing with limited data. Furthermore, the SiLU activation function is applied in the hidden layers, allowing smooth and nonlinear transformations that enhance both learning and reconstruction performance.

The method involves training an Auto-Encoder on a large dataset to establish a nonlinear mapping between the low-dimensional latent space and the high-dimensional data space. During reconstruction, sparse input data is used to compute the latent space coordinates, and the decoder then generates the complete reconstructed data field. The computational efficiency of this approach is attributed to the optimization process being primarily executed within the low-dimensional latent space.

Mathematically, an Auto-Encoder consists of an encoder \(\scaledEncoder: \mathbb{R}^{\nspacedof} \to \mathbb{R}^{\nbasisspace}\) and a decoder \(\scaledDecoder: \mathbb{R}^{\nbasisspace} \to \mathbb{R}^{\nspacedof}\). The encoder maps high-dimensional data \(\sol \in \mathbb{R}^{\nspacedof}\) to low-dimensional data \(\redsolapprox \in \mathbb{R}^{\nbasisspace}\), while the decoder reconstructs the data back to the high-dimensional space. The decoder serves as a nonlinear mapping \(\scaledDecoder(\redsolapprox)\), forming a nonlinear manifold.

The general structure of the encoder and decoder varies, but in this study, we employ the shallow and sparse network introduced in Section 3.2 of \cite{kim2022fast}. The encoder and decoder each consist of three layers and have a symmetric structure. The decoder's input and hidden layers are fully connected, with the SiLU activation function applied. Sparse connections without activation functions are used between the hidden and output layers. Two hyperparameters, \(b\) and \(\delta b\), define the sparse structure: $b$ represents the size of the previous layer node block used to compute a single output node, and $\delta b$ defines the distance between blocks. Further details can be found in \cite{kim2022fast}.

For model training, we use the snapshot matrix defined as:
\begin{equation}
    \snapshots\defeq\bmat{\solArg{0}^{\param_1} - \solArg{ref} &
\cdots & \solArg{\ntimedof}^{\param_{\nparam}}- \solArg{ref}}\in\RR{\ndof\times\nparam(\ntimedof+1)}, 
\end{equation}
where \(\solArg{n}^{\param_k}\) represents the solution at time step \(n\) and parameter \(\param_k\), with \(n \in \{0, \dots, \ntimedof\}\) and \(k \in \{1, \dots, \nparam\}\). For simplicity and readability, the notation $\solArg{n}^{\param_k}$ is used without explicitly indicating $n$ and $\param_k$ in the rest of this paper.

Gappy AE utilizes the nonlinear manifold defined as:
$\spatialSubspace \defeq {\scaledDecoder\left(\reddummy\right)|\reddummy \in \RR{\nbasisspace}}$. Here, \(\scaledDecoder: \mathbb{R}^{\nbasisspace} \to \mathbb{R}^{\nspacedof}\) maps the \(\nbasisspace\)-dimensional latent space to the \(\nspacedof\)-dimensional data space satisfying $\nbasisspace\ll\nspacedof$. The reconstruction is approximated as:
\begin{equation}\label{eq:spatialNMROMsolution}
    \sol \approx \solapprox = \solArg{ref} + \scaledDecoder(\redsolapprox),
\end{equation}
where \(\redsolapprox \in \mathbb{R}^{\nbasisspace}\) denotes the generalized coordinates. The initial estimate is given as \(\redsolapproxArg{0} = \scaledEncoder(\solArg{0} - \solArg{ref})\), where \(\scaledEncoder \approx \scaledDecoder^{-1}\) and $\solArg{0}$ is the initial value..

Since the measurement values used for data reconstruction are sparse, only a portion of the decoder's output is utilized. Because the decoder consists of sparse and shallow layers, the nodes involved in computing part of the output layer are also sparse. Therefore, the minimization problem for computing the generalized coordinates \(\hat{\boldsymbol{u}} \in \mathbb{R}^{\nbasisspace}\) in the reduced space, given by Equation \ref{eq:minimization}, can be solved efficiently using the Gauss-Newton method.
\begin{equation}\label{eq:minimization}
\hat{\boldsymbol{u}}=\argmin{\reddummy\in\RR{\nbasisspace}}\lVert \samplemat (\sol-\scaledDecoder(\reddummy)) \rVert_2^2,
\end{equation}
where $\samplemat\in\RR{\nbasisres\times\ndof}$ is a sampling matrix that extracts the rows corresponding to the measurement points from the solution matrix. $\samplemat$ is constructed as follows. First, define $\mathbf{n}\in \RR{\nspacedof}$ as:
\begin{equation}
n_j=\begin{cases}
1, & \text{if $x_j$ is known}\\
0, & \text{if $x_j$ is missing}
\end{cases}
\end{equation}
where $\sol \in \RR{\nspacedof}$ represents the solution vector for a given parameter, and $x_j$ is the $j$-th component of $\sol$. Next, construct the diagonal matrix $\mathbf{N}\in\RR{\nspacedof\times\nspacedof}$ with the elements of $\mathbf{n}$ along the diagonal and remove the zero columns to obtain $\mathbf{Z} \in \RR{\nspacedof\times\nbasisres}$. Here, $\nbasisres$ is the number of nonzero elements in vector $\mathbf{n}$, i.e., the number of measurement points. In the data reconstruction process (i.e., the online phase), $\samplemat \sol$ is replaced with sparse measurement data. Regardless of the measurement or observation locations, reconstruction can be performed without retraining the model, since the model is trained on the full dataset and does not depend on any specific measurement configuration.

\section{Sampling Algorithms}\label{sec:sampling}
The reconstruction performance of sparse data is heavily influenced by the location of the sampling points. To achieve the best possible reconstruction accuracy, it is essential to strategically determine the placement of these measurement points. This section explores the sampling algorithms applied in this study to optimize the performance of Gappy AE reconstruction method.

In this study, two sampling strategies were employed: Latin hypercube sampling (LHS) \cite{mckay2000comparison} and the Discrete Empirical Interpolation Method (DEIM) \cite{chaturantabut2010nonlinear, drmac2016new, drmac2018discrete, carlberg2013gnat, choi2020sns}. LHS is a near-random sampling method of parameter values from a multidimensional distribution. The fundamental principle of LHS is to divide the probability distribution of each variable into equal probability intervals. From each interval, a sample is then randomly selected. This approach ensures that the entire range of each variable is represented in the final sample set, providing a more comprehensive coverage of the parameter space compared to simple random sampling. 

The DEIM was introduced as a more advanced approach to strategically identify sampling points that minimize reconstruction errors. The DEIM algorithm operates by leveraging the Proper Orthogonal Decomposition (POD) basis matrix $\basismatspace=[\boldsymbol{\phi}_1,\cdots,\boldsymbol{\phi}_p] \in \RR{N\times p}$, where $\boldsymbol{\phi}_1, \cdots, \boldsymbol{\phi}_p$ are POD modes and represent the dominant modes of the field. This basis is derived from a set of high-fidelity snapshots. Using a greedy algorithm, DEIM selects sampling points that correspond to the locations with the highest absolute values in the POD modes. These points are chosen iteratively to ensure that the reconstruction error lower bound is minimized. This approach is particularly effective in scenarios where sparse data must represent the dominant characteristics of a high-dimensional field.

Unlike the linear subspace-based data reconstruction method, Gappy AE does not inherently rely on a basis matrix for such DEIM. To apply the DEIM algorithm to Gappy AE, following the approach introduced in \cite{KIM2024116978}, the residual between the original data and the reconstruction $\mathbf{r} \in \RR{\nspacedof}$ is defined as $\mathbf{r}=\sol - \scaledDecoder(\redsolapprox)$ and is approximated using a POD basis derived from residual snapshots as $\mathbf{r}\approx\Tilde{\mathbf{r}}=\basismatspace_r\hat{\mathbf{r}}$. Here, $\basismatspace_r \in \RR{\nspacedof\times\nbasisres}$ is the Proper Orthogonal Decomposition (POD) basis matrix obtained from the residual snapshot matrix, and $\hat{\mathbf{r}} \in \RR{\nbasisres}$ is the reduced residual. This residual POD basis matrix was used in the DEIM algorithm for Gappy AE.

In this study, both LHS and DEIM-based sampling were evaluated for reconstruction performance. It was observed that DEIM outperformed LHS, as the greedy selection process identified the most informative measurement locations.

\section{Numerical Experiments}\label{sec:numerical}
To assess the effectiveness of seismic data reconstruction using the Gappy AE algorithm, we conducted numerical experiments. The input data for reconstruction consists of ground motion data collected from sparse observation points, while the output data represents the reconstructed ground motion across the surrounding region. The data used for both measurement and reconstruction is the magnitude of horizontal ground motion velocity. To evaluate the performance of Gappy AE in data reconstruction, we used this quantity along with the Peak Ground Velocity (PGV)-based Modified Mercalli Intensity (MMI). For error computation, we calculated both projection error and reconstruction error. The projection error represents the reconstruction error when all data points are available, serving as the lower bound of Gappy AE’s reconstruction performance. The reconstruction error is defined as
\begin{equation}\label{eq:recon_error}
    \text{reconstruction error}(\%)=\frac{\lVert \sol-\solapprox \rVert_2^2}{\lVert \sol\rVert_2^2}\times100
\end{equation}
where $\sol$ denotes the ground truth seismic data, and $\solapprox$ represents the reconstructed one on the grid.

Details on the generation of synthetic data are provided in Section \ref{sec:offline}, while the reconstruction performance results are presented in Section \ref{sec:online}.

\subsection{Offline Phase}\label{sec:offline}
In Gappy data reconstruction, the offline phase refers to preprocessing steps for data restoration, including generating training data and training the model.

\begin{figure}[H]
  \centering
  \includegraphics[width=0.8\textwidth]{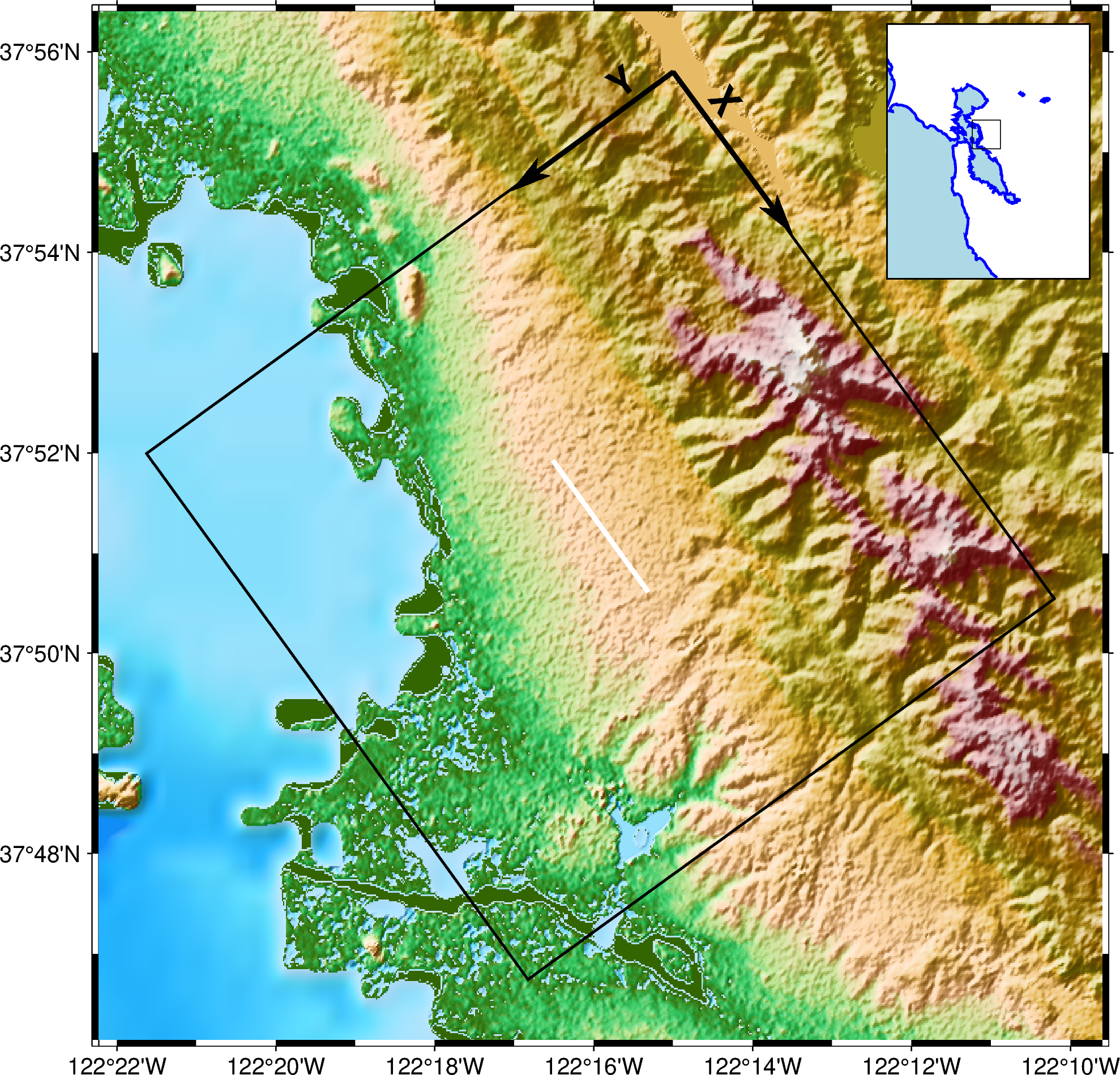}
  \caption{Simulation domain in the Berkeley area, outlined by the black rectangular box. Point-source earthquakes are positioned along the thick white line.}
  \label{fg:domain}
\end{figure}

Synthetic data was generated using SW4 \cite{sw4-v3.0, Sjogreen2012sw4}. A velocity model of the Berkeley area in California, USA was used, and topography was not considered. The size of the simulation domain is $(x,y,z) \in [0,12]\text{km}\times [0,12]\text{km} \times [0,6]\text{km}$, where the $+z$ direction points downward. The surface area of the domain is shown in Fig. \ref{fg:domain}. The data used for training and testing was sampled at uniform grid points with 100-meter intervals within the domain $(x,y) \in [3,9]\text{km}\times [3,9]\text{km}$ at time intervals of 8.69565e-3 seconds.

A focal mechanism considering the Hayward Fault in the Berkeley area was assumed, using a strike of 143 degrees, a rake of 0 degrees, and a dip of 90 degrees. The seismic source was modeled as a point source, and a Gaussian source time function with a 1-second offset and a frequency of 6.28 Hz was used. 

The simulation parameters included the epicenter location along the fault (i.e., x), the source depth (i.e., z), and the moment magnitude (i.e., m0). The parameter sweep is as follows:
\begin{itemize}
    \item x=\{4.5, 4.8, 5.1, 5.4, 5.7, 6. , 6.3, 6.6, 6.9, 7.2, 7.5\}km
    \item z=\{1.5, 1.7, 1.9, 2.1, 2.3, 2.5, 2.7, 2.9, 3.1, 3.3, 3.5\}km 
    \item m0=\{1, 2, 3, 4, 5, 6, 7, 8, 9, 10, 11\}$\times$1e16Nm
\end{itemize}
The data index $i$ corresponds to the combinations of x, z, and m0. The order follows a triple for-loop over x, z, and m0, incrementing by 1 at each step, where three loops iterate 11 times each. The index is updated according to $
i = 11^2 i_1 + 11 i_2 + i_3, \quad \text{where} \quad 0 \leq i_1, i_2, i_3 < 11.$ Each parameter combination corresponds to a unique index $i$ based on the values of $i_1,i_2,i_3$. The test data was selected from the following parameters:
\begin{itemize}
    \item x=\{4.5, 5.1, 5.7, 6.3, 6.9, 7.5\}km
    \item z=\{1.5, 1.9, 2.3, 2.7, 3.1, 3.5\}km 
    \item m0=\{1, 3, 5, 7, 9, 11\}$\times$1e16Nm
\end{itemize}

The autoencoder was trained using the data, excluding the test set. The data was split into train, validation, and test sets in a 0.67:0.17:0.16 ratio. 
Each simulation generated by SW4 initially contains 690 time steps, corresponding to 6 seconds of ground motion data after the onset of the earthquake. To reduce the data size while preserving the essential features, the time series was uniformly downsampled to 138 time steps, resulting in an effective sampling rate of approximately 23 Hz. The spatial domain is discretized with a $61 \times 61$ grid, covering a 6 km × 6 km area, which yields 3721 spatial points per time step. A total of 1115 different parameter configurations (e.g., source location, moment tensor) were sampled for training and validation. As a result, the complete dataset consists of $138 \times 1115 = 153\,870$ time instances, each represented by $3721$ spatial features. The data were stored in a 2D array of size $153\,870 \times 3721$, occupying approximately 2.13 GB. In this array, each row corresponds to a single time step from a specific simulation instance, and each column corresponds to a spatial grid point.

\begin{figure}[H]
  \centering
  \includegraphics[width=0.6\textwidth]{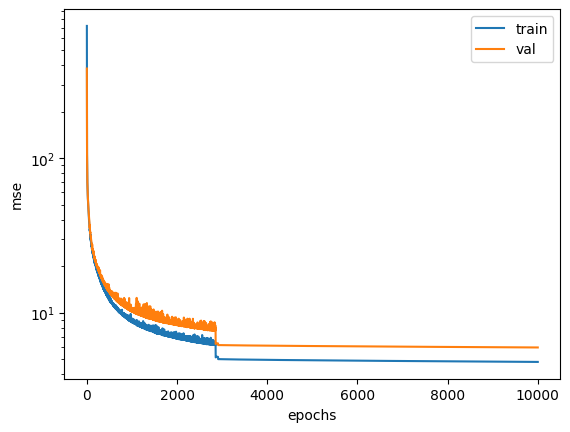}
  \caption{Autoencoder MSE Loss}
  \label{fg:ae_loss}
\end{figure}

The autoencoder used in this study has a symmetric structure between the encoder and decoder, with sparse and shallow layers. The autoencoder consists of three hidden layers, and the SiLU activation function was applied to all layers except the output layer. A mask matrix was applied between the hidden layers of the encoder and the bottleneck layer, as well as between the bottleneck layer and the decoder’s hidden layers, to enforce sparse connections. The size of the bottleneck layer is denoted as $ls$, representing the size of the latent space. The mask matrix is tuned using two hyperparameters, $b$ and $\delta b$, where $b$ is the block size, representing the number of nodes from the previous layer used to compute a single node value, and $\delta b$ is the gap between blocks. Thus, the hyperparameters determining the size of the sparse and shallow autoencoder used in this study are $b$, $\delta b$, and $ls$. The training and validation MSE losses for various model hyperparameters are shown in Table \ref{tab:exp_mseloss}.
\begin{table}[H]
    \centering
    \begin{tabular}{cccccc}
        \toprule
        $b$ & $\delta b$ & $ls$ & last epoch & train loss & val loss \\
        \midrule
        100  & 10  & 20 & 7638  & 11.04 & 11.87 \\
        200  & 20  & 20 & 7188  & 8.32  & 9.30  \\
        300  & 30  & 20 & 10\,323 & 6.78  & 7.86  \\
        400  & 40  & 20 & 20\,000 & 5.80  & 6.92  \\
        500  & 50  & 20 & 10\,000 & 5.16  & 6.30  \\
        600  & 60  & 20 & 10\,000 & 4.65  & 5.78  \\
        700  & 70  & 20 & 10\,000 & 4.28  & 5.42  \\
        1000 & 20  & 20 & 10\,000 & 4.80  & 5.95  \\
        1000 & 30  & 20 & 10\,000 & 5.05  & 6.24  \\
        1000 & 40  & 20 & 10\,000 & 4.77  & 5.97  \\
        1000 & 50  & 20 & 10\,000 & 4.11  & 5.25  \\
        1000 & 60  & 20 & 10\,000 & 4.36  & 5.54  \\
        1000 & 70  & 20 & 10\,000 & 3.94  & 5.08  \\
        100  & 10  & 25 & 7713  & 9.24  & 10.03 \\
        200  & 20  & 25 & 6835  & 6.93  & 7.85  \\
        300  & 30  & 25 & 8547  & 5.49  & 6.48  \\
        400  & 40  & 25 & 10\,000 & 4.63  & 5.63  \\
        100  & 10  & 30 & 8959  & 7.77  & 8.52  \\
        200  & 20  & 30 & 8236  & 5.58  & 6.44  \\
        300  & 30  & 30 & 8005  & 4.60  & 5.51  \\
        400  & 40  & 30 & 10\,000 & 4.23  & 5.18  \\
        \bottomrule
    \end{tabular}
    \caption{Mean Squared Error (MSE) Loss Based on Hyperparameter Configurations}
    \label{tab:exp_mseloss}
\end{table}
The epoch-wise MSE loss curve for the autoencoder, corresponding to the case with the lowest maximum reconstruction error in Gappy AE (as described in Section \ref{sec:offline}), where $ls=20$, $b=1000$, and $\delta b=20$, is shown in Fig. \ref{fg:ae_loss}. The sudden drop in loss around epoch $2850$ is due to a scheduled learning rate reduction from $1e^{-4}$ to $1e^{-6}$. This allowed the optimizer to make finer adjustments, leading to a sharp decrease in loss.

The ADAM optimization algorithm was used for training the autoencoders. The initial learning rate was set to $1e-4$, and the ReduceLROnPlateau learning rate scheduler was used with a patience of $50$ and a factor of $0.1$. The number of training epochs was set to $10\,000$ or $20\,000$, but training was terminated early if the mean squared error loss did not improve for 100 consecutive epochs or if the learning rate fell below $5e-7$. The autoencoder was trained on a single NVIDIA A100 GPU with 80GB of memory. The average training time per epoch for $ls=20$, $b=1000$, and $\delta b=20$ was $58.2358$ seconds. From this point forward, unless otherwise specified, 'the autoencoder' refers to the one with these hyperparameters.

\begin{figure}[H]
  \centering
  \includegraphics[width=0.6\textwidth]{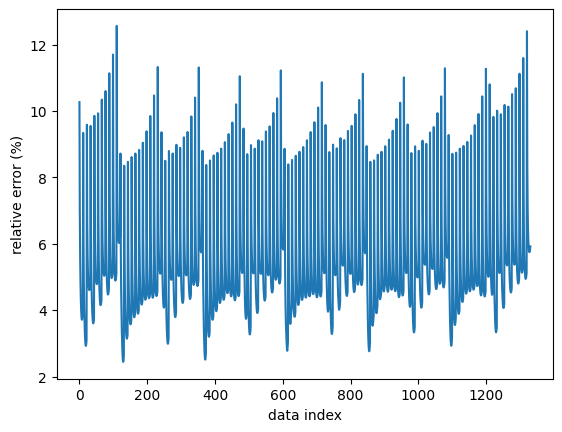}
  \caption{Gappy AE Projection Error (The data index is used to indicate each parameter combination.)}
  \label{fg:ae_projection_error}
\end{figure}

To examine the lower bound of the data reconstruction performance of the Gappy AE algorithm using this autoencoder, the projection error as defined in Section \ref{sec:numerical} was calculated and is shown in Fig. \ref{fg:ae_projection_error}. The projection error was computed using the entire dataset, including training, validation, and test data. Therefore, the range of parameter indices spans from 0 to 1330. The maximum error was $12.56\%$, the mean was $5.34\%$, and the minimum was $2.44\%$.

\subsection{Online Phase}\label{sec:online}
In Gappy data reconstruction, the online phase refers to the process of real-time data reconstruction based on sparse measurements. As mentioned in Section \ref{sec:Gappy AE}, the auto-encoder model training remains independent of changes in the  measurement locations.

First, we discuss the speed of data reconstruction. The computational cost of the online phase of Gappy data reconstruction is measured in terms of wall-clock time. Calculations are performed on a single CPU of an AMD EPYC 7763 @ 2.45 GHz with DDR4 Memory @ 3200 MT/s. With $61$ DEIM sample points and a reduced dimension (i.e., the latent space dimension) of $20$, the online phase of the Gappy AE method takes $6.63$ ms per reconstruction step in our example. This corresponds to a processing speed capable of handling measurement data at $150$ Hz. Here, the term \textit{per reconstruction step} refers to the process of reconstructing the complete spatial field of ground motion at a single time step, based on sparse measurements from a subset of grid points. In other words, given sparse measurements (e.g., from 61 grid points) at a specific time, the Gappy AE algorithm reconstructs the ground motion at all grid points for that time step. Thus, the reported value of $6.63$~ms indicates the time required to perform the spatial reconstruction for one time step.

Next, we discuss data reconstruction errors. The performance of data reconstruction was evaluated based on the three hyperparameters that determine the size of the autoencoder model, as well as the LHS and DEIM sampling algorithms described in Section~\ref{sec:sampling}. The number of measurement points, which refer to the sparse observation locations used as input to the Gappy AE algorithm, was varied between $40$ and $61$ to compare reconstruction performance. These sampled points are selected from the $61 \times 61 = 3721$ spatial grid points and represent the subset of locations where ground motion is assumed to be measured. Using these sparse measurements at a given time step, the Gappy AE algorithm reconstructs the ground motion field over the entire grid (i.e., all 3721 spatial points) for that time step. In Table~\ref{tab:ae_recon_error}, for the LHS sampling algorithm, only cases where the maximum relative error was below $17.247\%$ are shown. For the DEIM sampling algorithm, only cases where the maximum relative error was below $15.616\%$ are included.
\begin{table}[H]
    \centering
    \resizebox{\textwidth}{!}{
    \begin{tabular}{cccccc}
        \toprule
        ($b$, $\delta b$, $ls$) & Sample Alg. & Num. Samples & Max. Rel. Err. (\%) & Mean. Rel. Err. (\%) & Min. Rel. Err. (\%) \\
        \midrule
        (1000, 10, 20) & LHS  & 57 & 17.247 & 8.285 & 4.939 \\
        (1000, 20, 20) & LHS  & 60 & 16.910 & 7.621 & 4.261 \\
        (1000, 30, 20) & LHS  & 59 & 17.222 & 7.553 & 3.616 \\
        (1000, 10, 20) & DEIM & 61 & 15.616 & 7.250 & 4.569 \\
        (1000, 20, 20) & DEIM & 60 & 15.520 & 6.653 & 3.779 \\
        (1000, 20, 20) & DEIM & 61 & 15.400 & 6.632 & 3.767 \\
        \bottomrule
    \end{tabular}
    }
    \caption{Relative Errors Based on Model Parameters and Sample Points}
    \label{tab:ae_recon_error}
\end{table}

\begin{figure}[H]
  \centering
  \subfigure[LHS]{
  \includegraphics[width=0.45\textwidth]{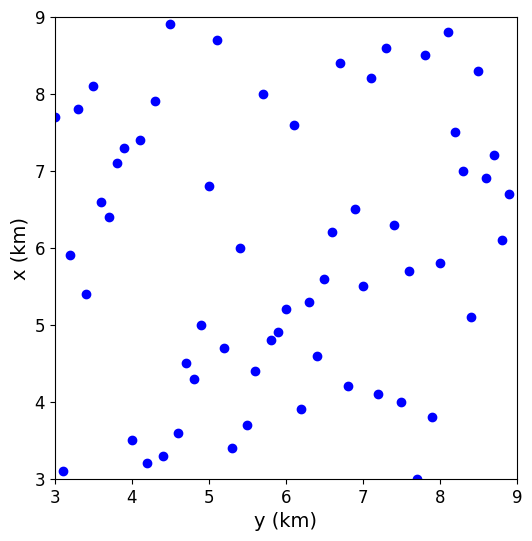}}
  \subfigure[DEIM]{
  \includegraphics[width=0.45\textwidth]{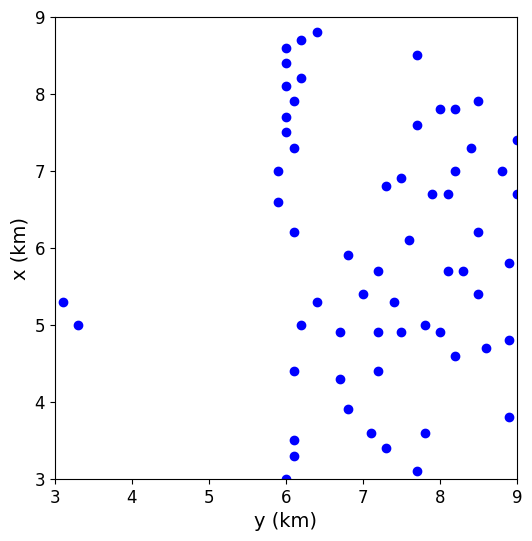}}
  \caption{Sampling Points}
  \label{fg:ae_sampling}
\end{figure}

\begin{figure}[H]
  \centering
  \includegraphics[width=0.6\textwidth]{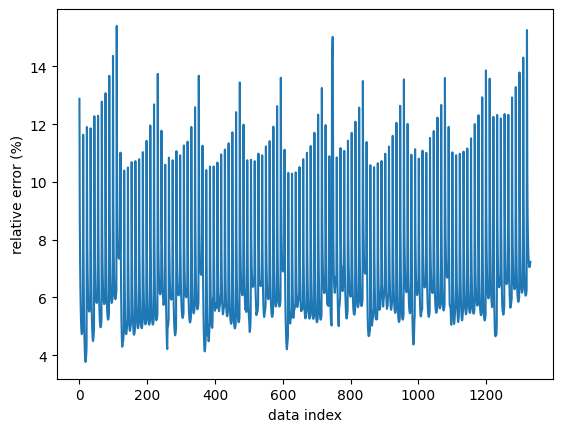}
  \caption{Gappy AE Reconstruction Error}
  \label{fg:ae_recon}
\end{figure}

\begin{figure}[H]
  \centering
  \subfigure[Ground Truth]{
  \includegraphics[width=1.0\textwidth]{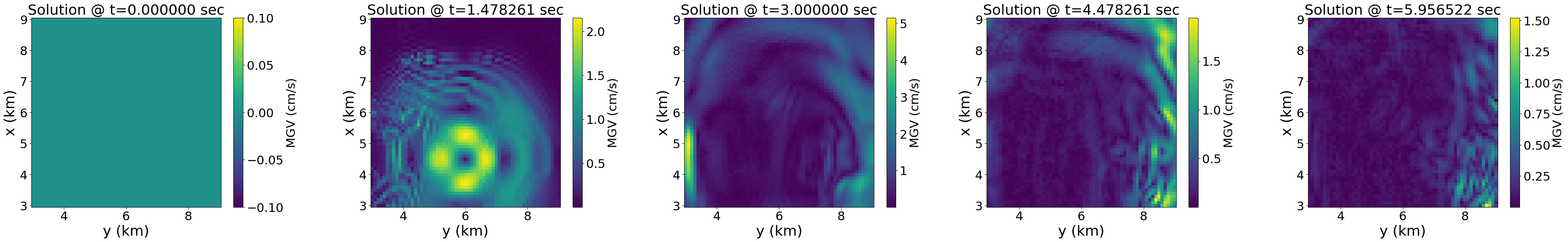}}
  \subfigure[Reconstruction]{
  \includegraphics[width=1.0\textwidth]{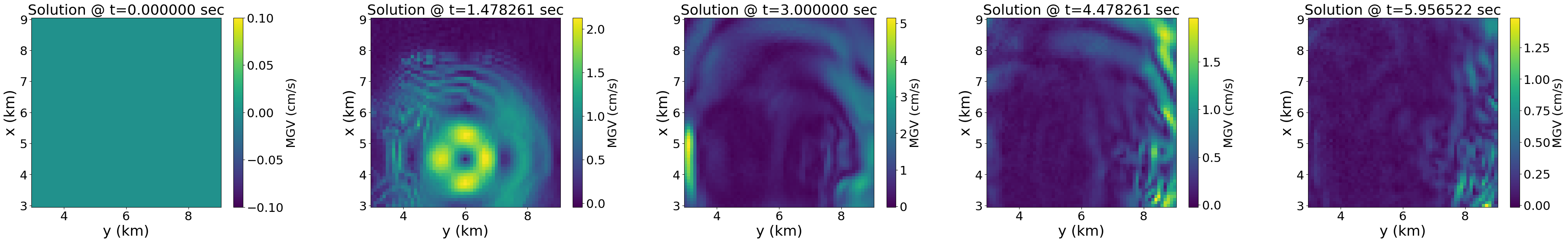}}
  \caption{History of MGV. Best Case (x=4.5km, z=1.7km, m0=8e16Nm)}
  \label{fg:ae_best_recon_mgv}
\end{figure}

\begin{figure}[H]
  \centering
  \subfigure[Ground Truth]{
  \includegraphics[width=1.0\textwidth]{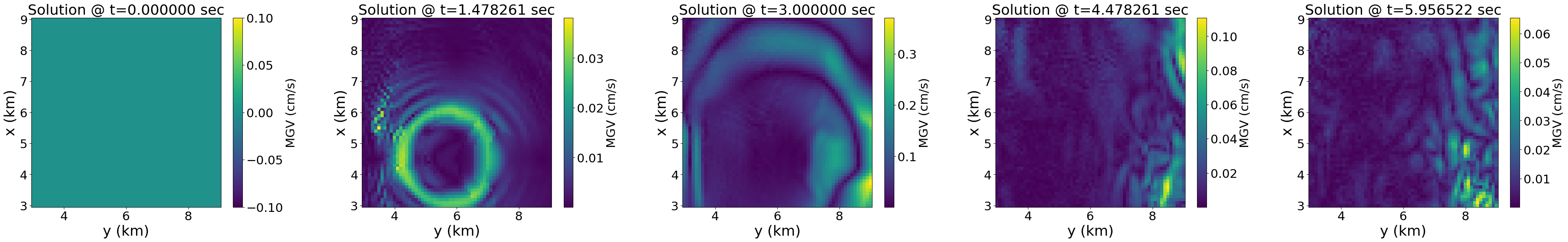}}
  \subfigure[Reconstruction]{
  \includegraphics[width=1.0\textwidth]{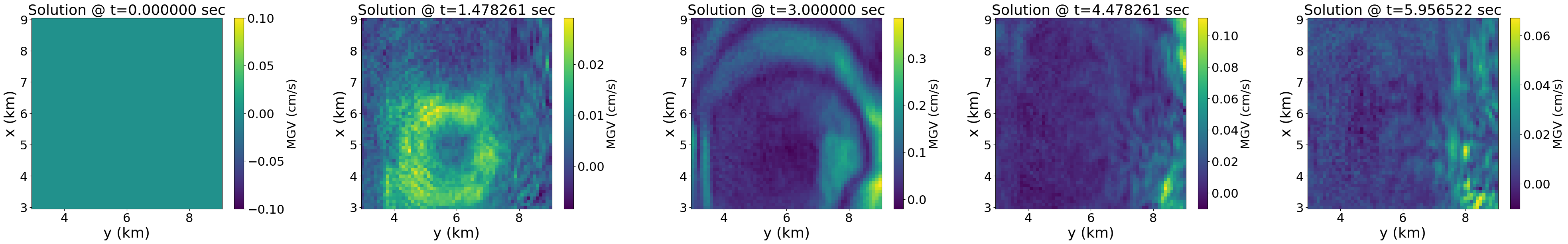}}
  \caption{History of MGV. Worst Case (x=4.5km, z=3.5km, m0=1e16Nm)}
  \label{fg:ae_wrost_recon_mgv}
\end{figure}

The results with the best reconstruction performance are presented using the Magnitude of Ground Velocity (MGV) and Modified Mercalli Intensity (MMI). Here, MGV refers to the magnitude of horizontal ground motion velocity, defined as $\sqrt{v_x^2 + v_y^2}$, where $v_x$ and $v_y$ are the horizontal components of ground motion velocity. The autoencoder model with $b=1000$, $\delta b=20$, and $ls=20$ showed the best performance in terms of maximum relative error for each sampling algorithm when using $60$ LHS measurement points and $61$ DEIM measurement points. These sampling points are shown in Fig. \ref{fg:ae_sampling}.
Between the two cases, the best reconstruction performance in terms of maximum relative error was achieved using $61$ DEIM sampling points. The MGV reconstruction errors of Gappy AE for all parameters are shown in Fig. \ref{fg:ae_recon}. As explained in Section \ref{sec:offline}, each data index corresponds to a combination of the simulation parameters (i.e., x, y, and m0).
The maximum, mean, and minimum reconstruction errors based on parameters were $15.40\%$, $6.63\%$, and $3.77\%$, respectively. The MGV plots with the smallest and largest reconstruction errors are shown in Fig. \ref{fg:ae_best_recon_mgv} and Fig. \ref{fg:ae_wrost_recon_mgv}, respectively.

\begin{figure}[H]
  \centering
  \includegraphics[width=0.6\textwidth]{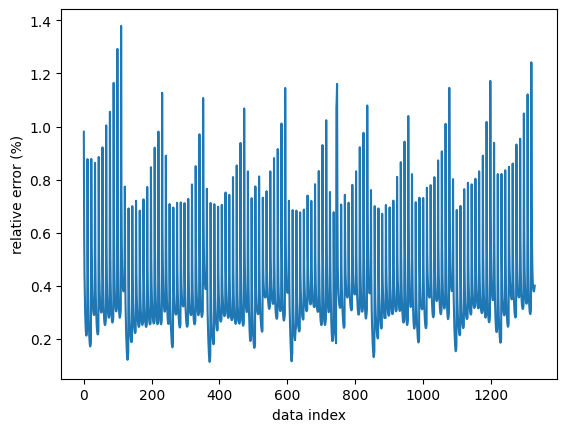}
  \caption{MMI Relative Error}
  \label{fg:MMI_error}
\end{figure}

\begin{figure}[H]
  \centering
  \includegraphics[width=1.0\textwidth]{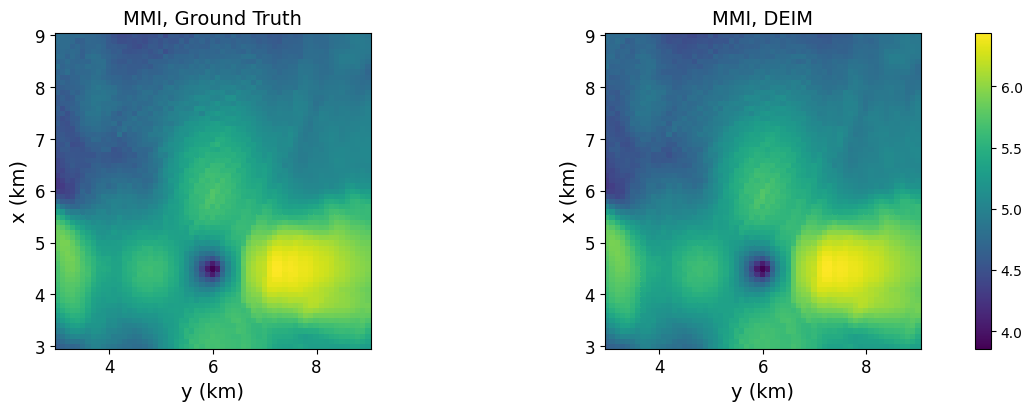}
  \caption{MMI. Best Case (x=4.5km, z=1.7km, m0=8e16Nm)}
  \label{fg:ae_best_recon_mmi}
\end{figure}

\begin{figure}[H]
  \centering
  \includegraphics[width=1.0\textwidth]{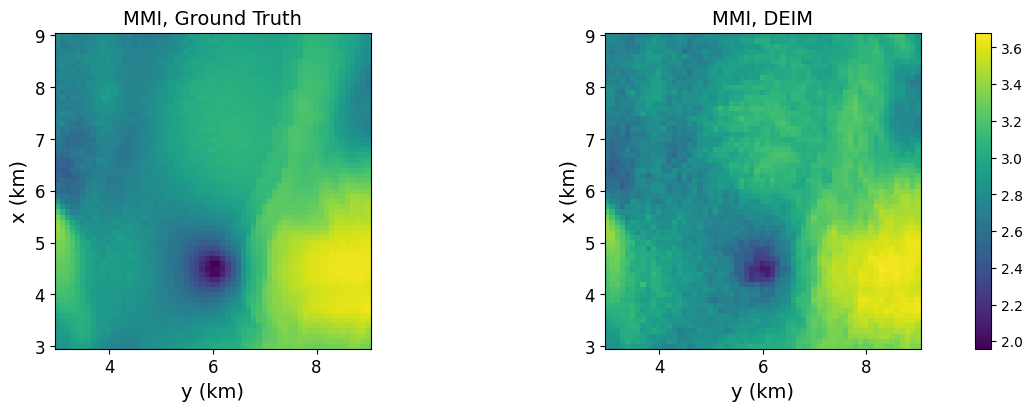}
  \caption{MMI. Worst Case (x=4.5km, z=3.5km, m0=1e16Nm)}
  \label{fg:ae_wrost_recon_mmi}
\end{figure}

The real-time MGV values during the earthquake were reconstructed, and the PGV-based MMI was calculated and compared with the ground truth. The relative errors in MMI for all parameters are presented in Fig. \ref{fg:MMI_error}, with a mean relative error of $0.38\%$. The case with the smallest relative error is shown in Fig. \ref{fg:ae_best_recon_mmi}, with a relative error of $0.175\%$. The case with the largest relative error is shown in Fig. \ref{fg:ae_wrost_recon_mmi}, with a relative error of $1.379\%$. Since MMI calculations use the logarithmic value of PGV, the relative error values are smaller than those of MGV.

Using simulation data, we assumed a scenario with sparse measurements, reconstructed the MGV, and computed the MMI accordingly. The numerical results indicate that Gappy AE successfully captured the data patterns and achieved first-order accuracy.

\section{Field Data}

\subsection{Berkeley Seismic Event}
At 10:39:37 UTC on January 4, 2018, a magnitude 4.4 earthquake occurred 2 km southeast of Berkeley, California, USA. The epicenter of this earthquake was measured at latitude 37.855°N and longitude 122.257°W, with a focal depth of 12.3 km. The earthquake was recorded by the Northern California Seismic System (NC), a part of the California Integrated Seismic Network. According to data from the United States Geological Survey (USGS), a total of 289 seismic stations analyzed 324 waveforms to determine the location and magnitude of the event. Detailed information is available at \href{https://earthquake.usgs.gov/earthquakes/eventpage/nc72948801/origin/detail}{USGS}. The moment magnitude (Mw) of the earthquake was calculated as 4.37, and the released elastic energy was estimated to be $4.568 \times 10^{15}$ N-m. Considering the location of the epicenter and fault movement characteristics, this earthquake is  associated with the Hayward Fault. Moment tensor analysis indicates that the earthquake's fault movement follows a typical strike-slip fault mechanism. The two nodal planes analyzed were: (1) strike 59°, dip 87°, rake -13°, and (2) strike 150°, dip 77°, rake -177°. These characteristics align with the faulting style of Northern California and suggest that accumulated stress along the Hayward Fault was released in this event. Earthquakes of this magnitude occur relatively frequently in the Berkeley and San Francisco Bay Area \cite{USGS}.

\begin{figure}[H]
  \centering
  \includegraphics[width=0.8\textwidth]{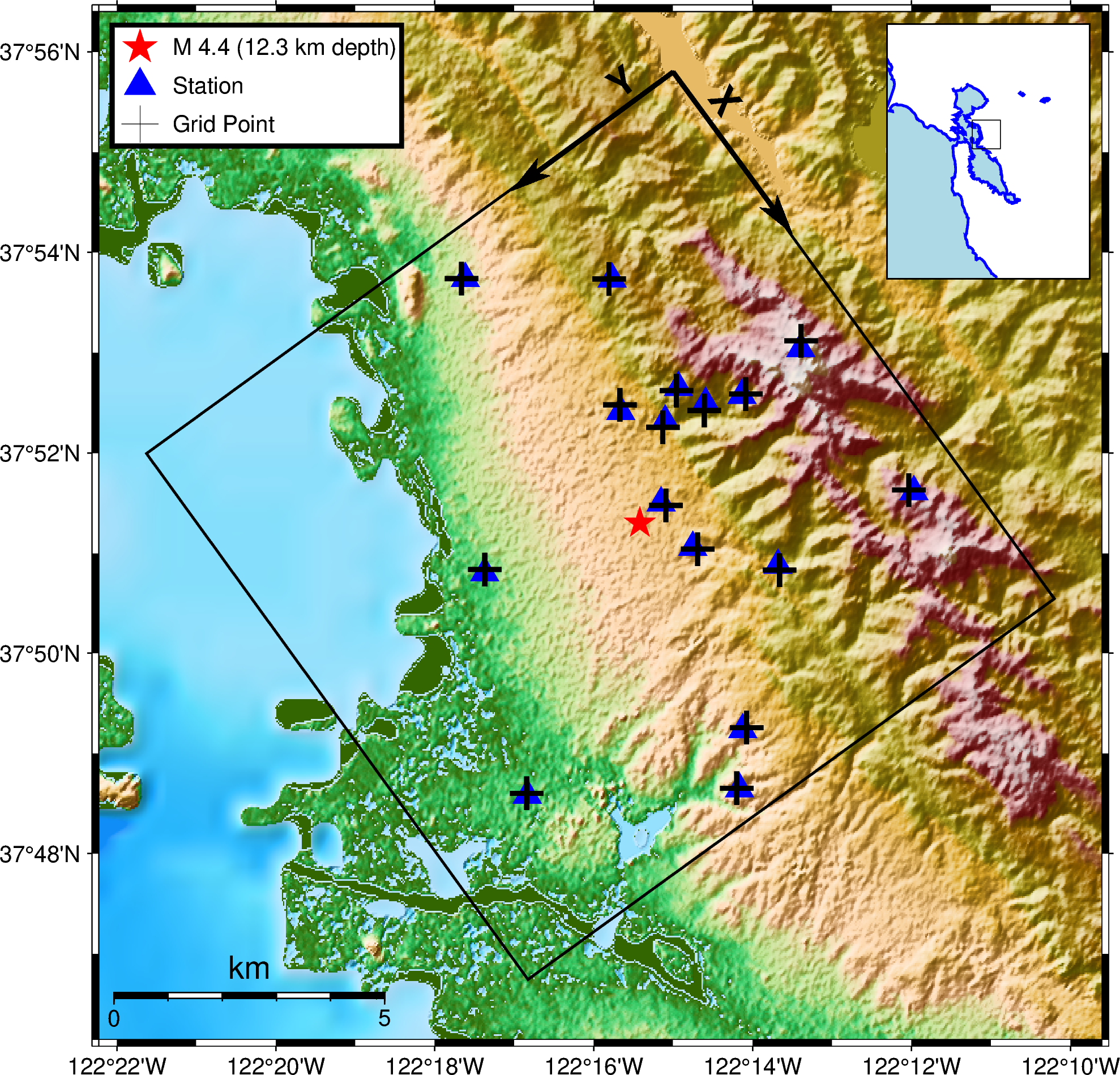}
  \caption{Map of the Berkeley area showing the locations of the stations and the epicenter of the Berkeley event.}
  \label{fg:sta}
\end{figure}

The epicenter of this earthquake near Berkeley, along with the locations of observation stations in a 12 km × 12 km area around the city and topographic information, is shown in Fig. \ref{fg:sta}. In the figure, blue $\times$ markers represent the actual locations of observation stations, while black $+$ markers indicate the nearest grid points to these stations. The measurement data used in the data reconstruction algorithm is assumed to be collected at these grid points. Additionally, the data was resampled to 40 Hz and filtered between 0.1–20 Hz before downsampling to prevent aliasing.

\subsection{Offline Phase}
The number of observation stations within the reconstruction domain used in Section \ref{sec:offline} is nine. In this case, the latent space dimension of the autoencoder for data reconstruction should be less than half of nine to avoid numerical instability when solving the nonlinear problem in the Gappy AE algorithm. However, since the parameter dimension of the training data used for the autoencoder is three, and considering the time dimension, the intrinsic dimension of the training data is four. This implies that the latent space of the autoencoder should be greater than four to effectively learn data representation. In other words, due to limitations in data representation learning, additional observation stations were required. To address this, the data domain for autoencoder training was expanded to 12 km × 12 km, increasing the number of available observation stations to 16. It is noted that the same SW4 simulation data and parameter space were used as in Section~\ref{sec:offline}. While the training in the Numerical Experiments' offline phase was based on a $6\text{km} \times 6\text{km}$ subregion of the full simulation domain, the autoencoder in this section was trained using data from the entire $12\text{km} \times 12\text{km}$ surface area of the simulation output.

The autoencoder training process follows the same procedure as the offline phase in Section \ref{sec:offline} of Numerical Experiments. The hyperparameters defining the autoencoder structure were set to $b = 1000$, $\delta b = 20$, $ls = 6$. The train loss and validation loss of the autoencoder were 7.637 and 8.322, respectively. The projection error of the Gappy AE had a maximum value of $18.023\%$, an average value of $7.642\%$, and a minimum value of $5.307\%$.

\subsection{Online Phase}
When measurement values are provided in real time, the data reconstruction process is performed simultaneously in real time. The computational cost of the online phase of Gappy data reconstruction is measured in terms of wall-clock time. Calculations are conducted on a single CPU of an AMD EPYC 7763 @ 2.45 GHz with DDR4 memory @ 3200 MT/s. With $15$ sample points and a reduced dimension of $6$, the online phase of the Gappy AE method takes $34.04$ ms per reconstruction step in this example, enabling a processing speed of up to $29.38$ Hz.

As shown in Fig. \ref{fg:sta}, there are 16 observation stations available. To evaluate the performance of the Gappy AE algorithm, we select data from one station for validation and use data from the remaining 15 stations for reconstruction. MGV was compared between the reconstructed and recorded values in the time domain, the frequency domain using Fourier Amplitude, and in terms of Duration time, as used in \cite{Arben2024}. On top of that, we calculate MMI based on the reconstructed MGV data and compares it with the USGS ShakeMap.
Our reconstruction performance target for real field data is first-order accuracy. In other words, it refers to how well the pattern matches when plotting the MGV over time at a given observation station. 

\begin{figure}[H]
  \centering
  \includegraphics[width=1.0\textwidth]{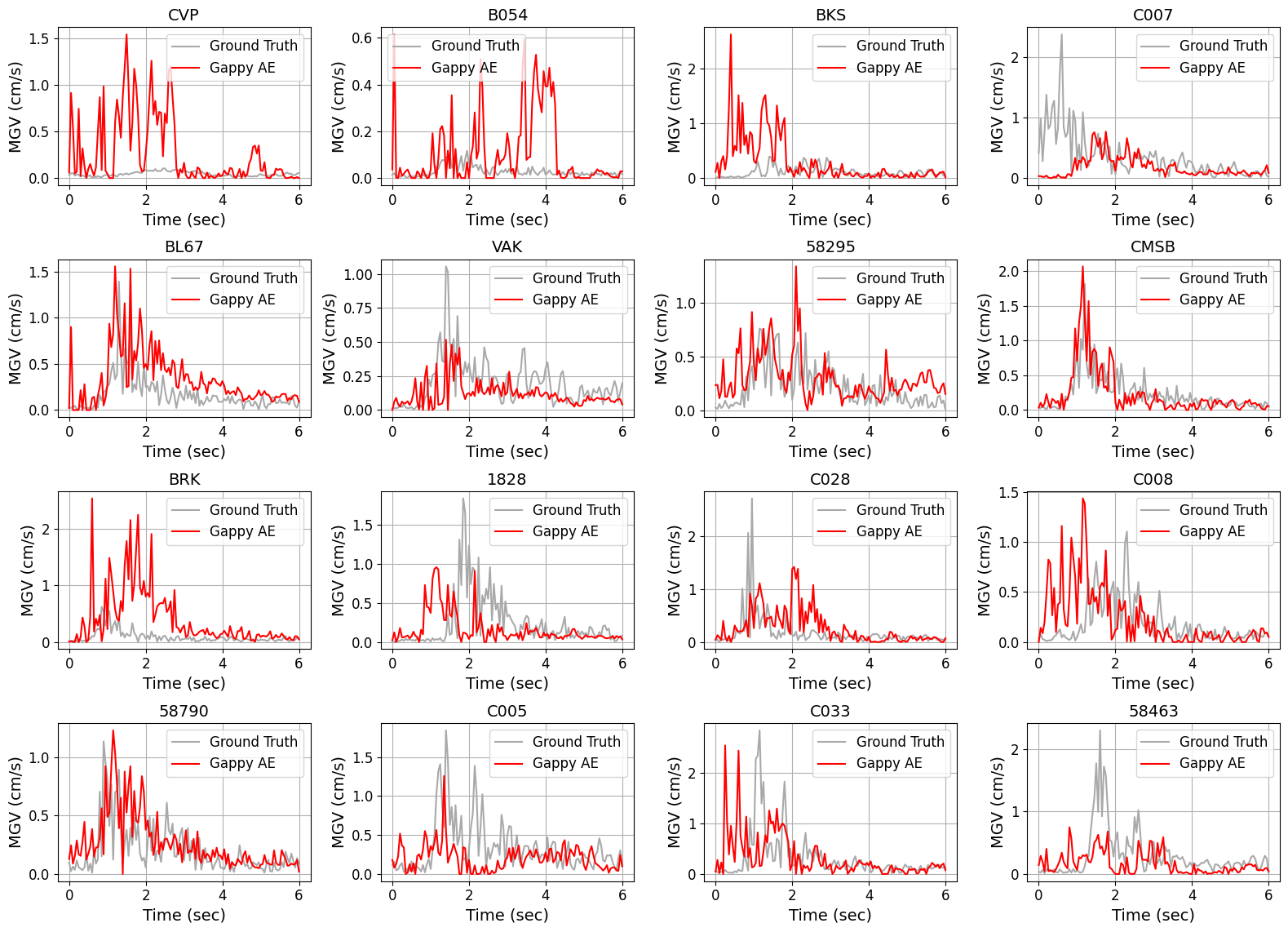}
  \caption{Comparison of reconstructed (red) and recorded (grey) time histories of MGV}
  \label{fg:field_mgv_time}
\end{figure}

In Fig. \ref{fg:field_mgv_time}, the MGV reconstruction data and measured (or recorded) data for all 16 stations are shown. The reconstructed data for each station was obtained using observation data from the remaining 15 stations. The curves of the recorded and reconstructed data for the 58295, CMSB, and 58790 stations showed a similar shape, whereas the others exhibited noticeable differences. This confirms that the reconstruction performance varies significantly depending on the location of the measurement points.

\begin{figure}[H]
  \centering
  \includegraphics[width=1.0\textwidth]{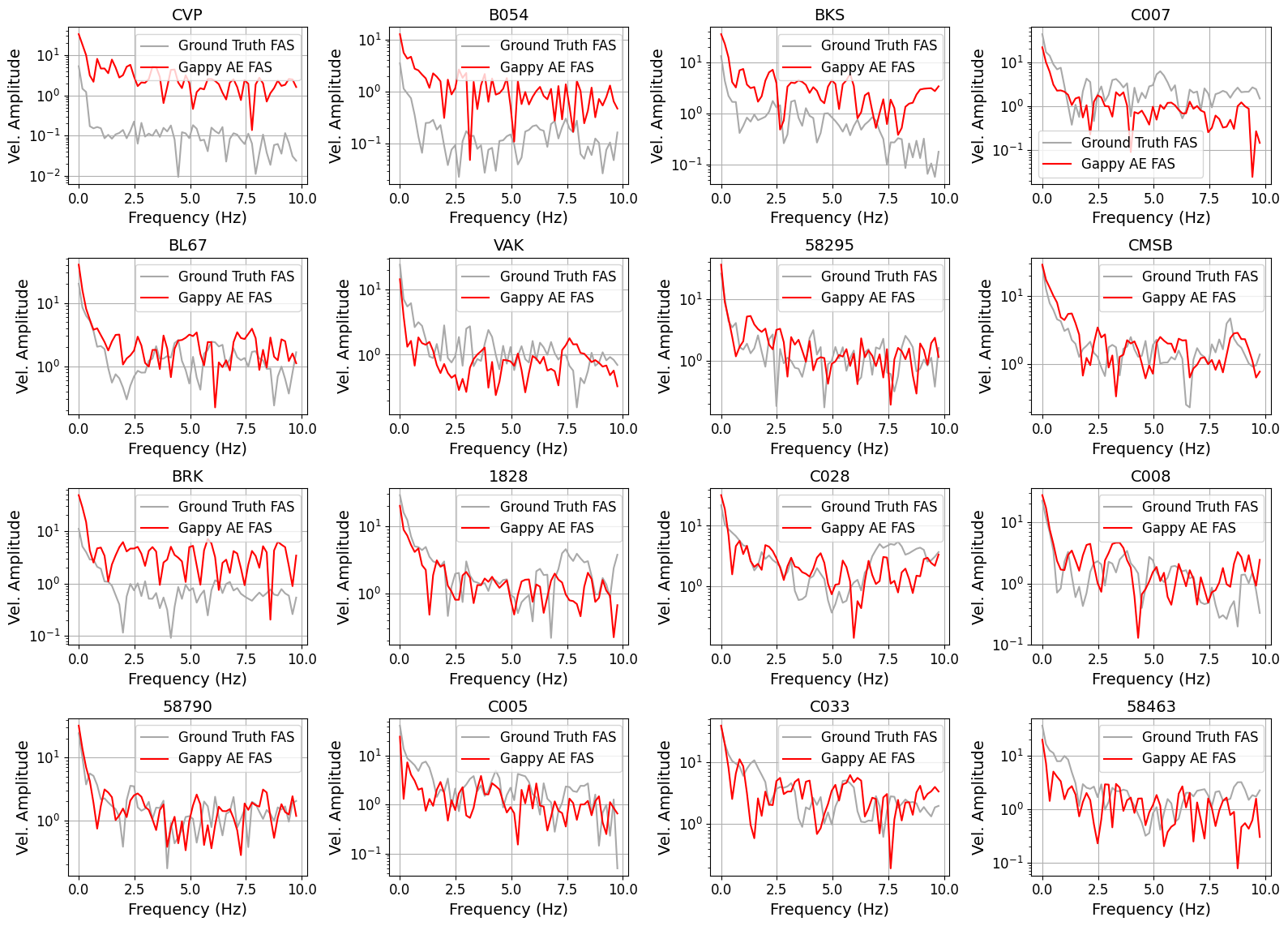}
  \caption{Comparison of reconstructed (red) and recorded (grey) Fourier amplitude spectra of ground motion}
  \label{fg:field_mgv_freq}
\end{figure}

\begin{figure}[H]
  \centering
  \includegraphics[width=1.0\textwidth]{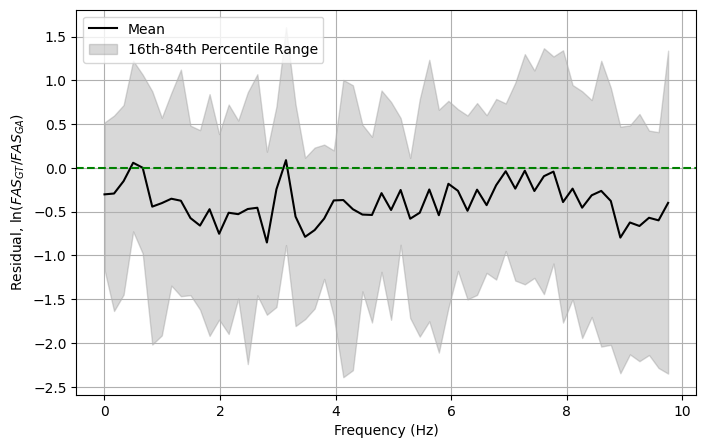}
  \caption{Comparison of the station-averaged Fourier amplitude residuals. The solid line shows the natural logarithm of the amplitude ratio between recorded and reconstructed motions across different frequencies, while the shaded region represents the 16th–84th percentile range of the residuals.}
  \label{fg:field_amp_res}
\end{figure}

The difference between the reconstructed and recorded data was analyzed in the frequency domain by comparing the Fourier Amplitude Spectra in Fig. \ref{fg:field_mgv_freq}. When examining the residuals on a natural log scale as shown in Fig. \ref{fg:field_amp_res}, a slightly larger discrepancy between the reconstructed and measured values was observed at higher frequencies. Additionally, across most frequency ranges, the reconstructed values generally exhibited lower amplitudes on average.

\begin{figure}[H]
  \centering
  \includegraphics[width=1.0\textwidth]{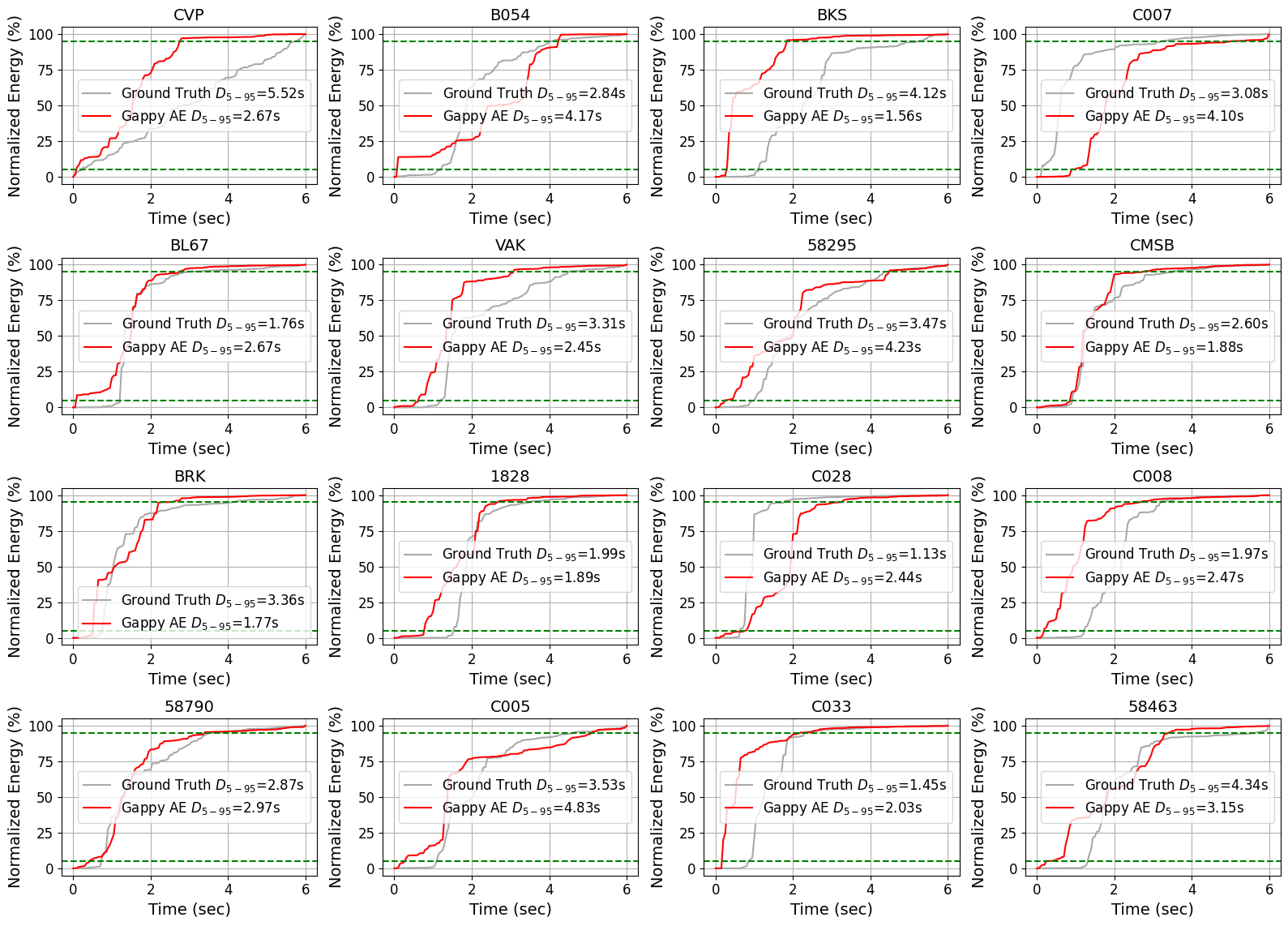}
  \caption{Comparison of reconstructed (red) and recorded (grey) normalized Aria's intensity. $D_{5-95}$ duration times are shown in each plot's legend.}
  \label{fg:field_duration}
\end{figure}

In Fig. \ref{fg:field_duration}, the normalized Aria’s intensity was compared to analyze how seismic energy spreads over time, focusing on the duration time between $5\%$ and $95\%$. According to the duration time analysis of the Berkeley earthquake in \cite{Arben2024}, most recorded data had longer duration times than the simulation data. However, when using reconstructed data, the number of cases where the duration time was longer or shorter than the recorded data was roughly equal. At the CVP, BKS, VAK, CMSB, BRK, 1828, and 58463 stations, the recorded data showed longer duration times compared to the reconstructed data, while in the remaining nine cases, the recorded data had shorter duration times. These time differences stem from variations in reconstruction accuracy and discrepancies between the velocity model used to generate the training data and the actual geological structure. In complex geological structures, wave scattering occurs, generating secondary coda waves after the primary shear waves, leading to longer durations. Therefore, if the reconstruction performance is sufficiently accurate, the difference in duration time could provide insights into discrepancies between the velocity model used in simulations and the real geological structure. However, the results presented in this section indicate that cases where the recorded data had longer and shorter duration times are nearly evenly distributed, suggesting that the reconstruction performance is not sufficiently accurate.

\begin{figure}[H]
  \centering
  \subfigure[MMI from reconstructed MGV]{
  \includegraphics[width=0.45\textwidth]{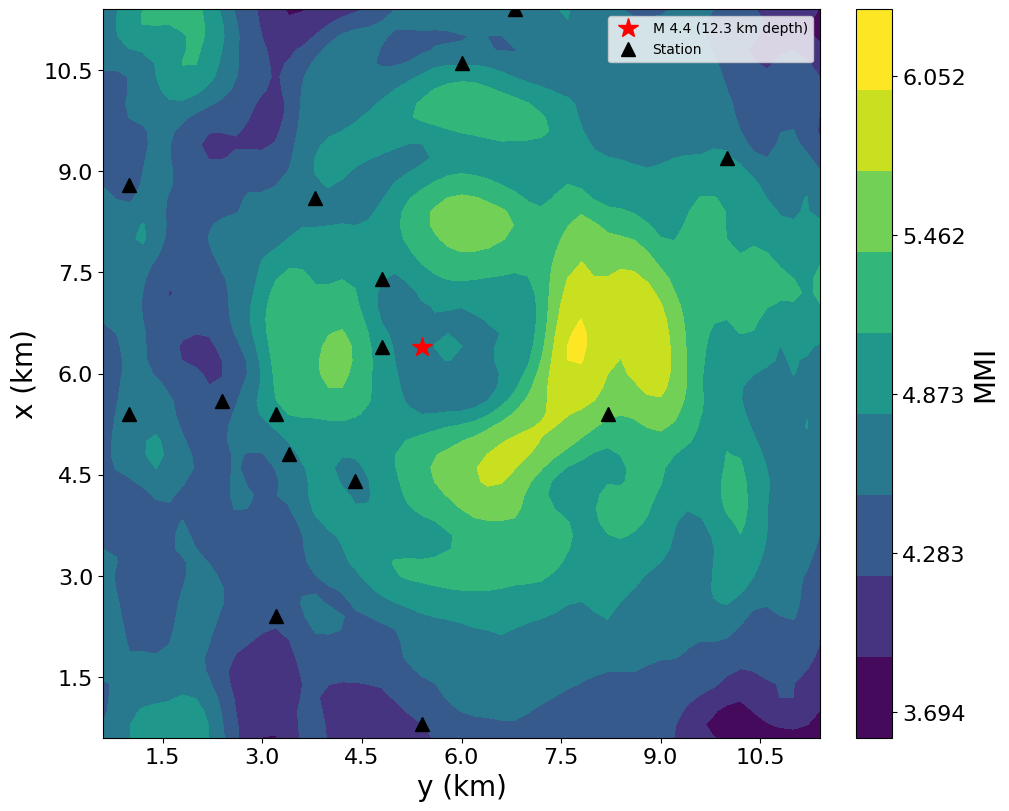}}
  \subfigure[USGS ShakeMap]{
  \includegraphics[width=0.45\textwidth]{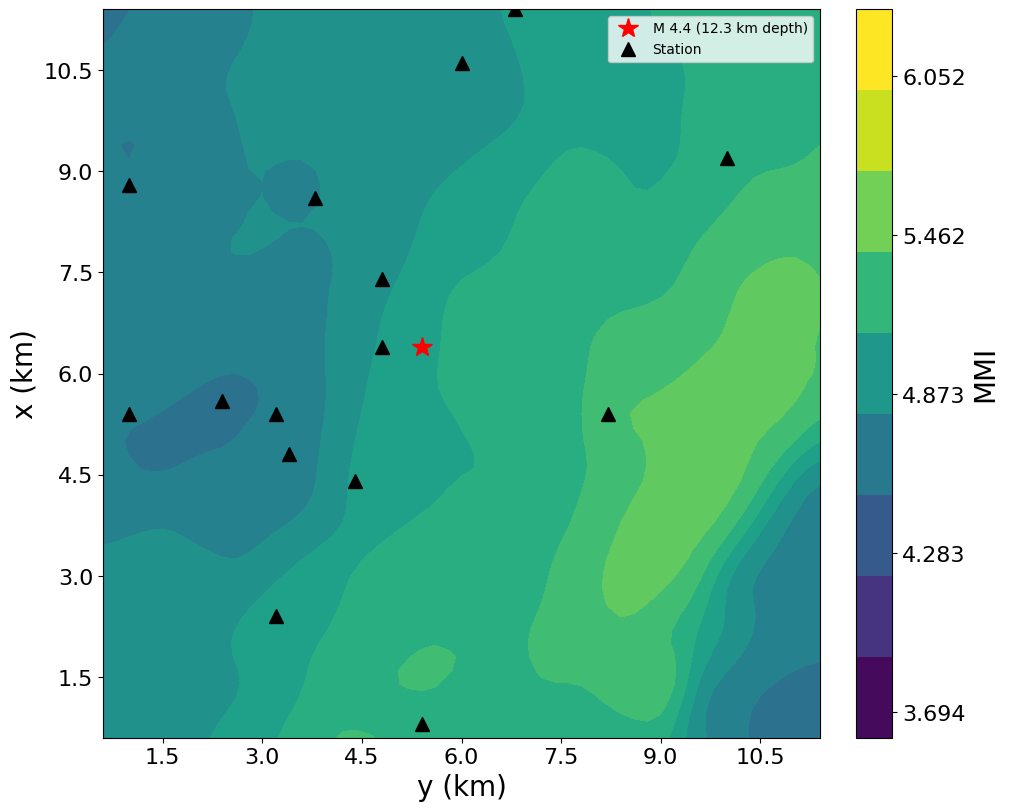}}
  \caption{Comparison of MMI from reconstructed MGV (a) and USGS ShakeMap (b). The triangular symbols represent the observation stations whose measurements were used in the reconstruction process.}
  \label{fg:field_MMI}
\end{figure}

In Fig. \ref{fg:field_MMI}, a comparison was made between the MMI plot calculated using the reconstructed MGV from 15 station measurements (excluding CMSB) near the Berkeley area and the ShakeMap provided by USGS. While the detailed distribution of MMI did not perfectly match, a trend was observed where the western part of the epicentral region (i.e., the right side of the x-y plane) exhibited higher MMI values compared to the eastern part (i.e., the left side of the x-y plane).

The source parameters of the simulation data used to train the auto-encoder that is used for Gappy AE fell outside the parameter range of the 2018 M4.4 Berkeley event. This means that the field data test evaluated the generalization capability of Gappy AE, which resulted in lower performance compared to its accuracy on synthetic data as shown in Section \ref{sec:online}. If the training dataset is augmented with parameters that include source characteristics of seismic events frequently occurring in the Berkeley area, the reconstruction performance on field data is expected to improve and align more closely with the high accuracy observed in synthetic data experiments. We believe that the results shown here highlights the potential applicability of the Gappy AE algorithm to seismic data, which is the objective of this article.

\section{Conclusion}
This study proposed a real-time ground motion reconstruction method with sparse sensor data using the Gappy AE algorithm. Numerical experiments validated the method's efficiency and accuracy. Additionally, field data was used for reconstruction, demonstrating the practical applicability of the Gappy AE algorithm.

One of the key advantages of the Gappy AE method is its real-time applicability. By eliminating the need for high-fidelity simulations, the approach allows for efficient and fast data reconstruction. Additionally, it achieves high reconstruction accuracy even when using a minimal number of sensors, making it a practical solution in scenarios where measurement points are limited.

To enable real-time deployment, the Gappy AE algorithm can be triggered immediately as soon as ground motion measurements arrive from seismic stations. This allows the system to continuously reconstruct the full spatial distribution of ground motion at each time step with minimal delay. As a result, it becomes possible to obtain spatially dense ground motion information based on sparse sensors. This feature makes the Gappy AE approach especially useful for rapid situation assessment and early response in earthquake early warning systems or disaster management scenarios.

However, certain limitations remain. The computational cost during the training phase is still high, which could pose challenges for large-scale applications. Furthermore, the study did not include specific tests to assess the method's sensitivity to noise in measurements. Addressing this limitation is crucial for ensuring the robustness of the reconstruction process in real-world environments.

Future research will focus on training a noise-robust model. One approach under consideration is augmenting the training dataset with artificially added noise, allowing the model to learn to handle varying levels of measurement uncertainty. Additionally, efforts will be made to scale up the method from local regions to larger regional areas by leveraging an extensive seismic observation network, thereby enhancing the reconstruction of earthquake data across broader regions.

\section*{Acknowledgments}
This research was performed at the Korea Institute of Science and Technology and was supported by the Basic Science Research Program through the National Research Foundation of Korea (NRF), funded by the Ministry of Education (RS-2023-00272582). This research was also supported by the Ministry of Trade, Industry, and Energy and the Korea Evaluation Institute of Industrial Technology research grant (20012462). Y. Choi acknowledges partial support from the U.S. Department of Energy, Office of Science, Office of Advanced Scientific Computing Research, as part of the CHaRMNET Mathematical Multifaceted Integrated Capability Center (MMICC) program, under Award Number DE-SC0023164 at Lawrence Livermore National Laboratory. Q. Kong and Y. Choi also acknowledge support from Lab Directed Research and Development project at Lawrence Livermore National Laboratory, 24-ERD-012. Lawrence Livermore National Laboratory is operated by Lawrence Livermore National Security, LLC, for the U.S. Department of Energy, National Nuclear Security Administration under Contract DE-AC52-07NA27344. LLNL release number: LLNL-JRNL-2004860

\bibliographystyle{plain}
\bibliography{references}  

\end{document}